\RequirePackage{fix-cm}
\documentclass[smallextended]{svjour3}       
\smartqed  
\usepackage{graphicx}
\usepackage{amsmath}
\usepackage{amssymb}
\usepackage{marvosym}
\usepackage[numbers]{natbib}
\journalname{}
\begin{document}

\title{Cross-media Similarity Metric Learning with Unified Deep Networks
}


\author{Jinwei Qi$^1$   \and
        Xin Huang$^1$   \and
        Yuxin Peng$^1$
}

\authorrunning{Jinwei Qi et al.} 

\institute{
	\Letter  Yuxin Peng \at
              \email{pengyuxin@pku.edu.cn} \\
              \\
    1         Institute of Computer Science and Technology, Peking University, Beijing 100871, China\\
}

\date{Received: date / Accepted: date}

\maketitle

\begin{abstract}
As a highlighting research topic in the multimedia area, cross-media retrieval aims to capture the complex correlations among multiple media types. Learning better shared representation and distance metric for multimedia data is important to boost the cross-media retrieval. Motivated by the strong ability of deep neural network in feature representation and comparison functions learning, we propose the Unified Network for Cross-media Similarity Metric (UNCSM) to associate cross-media shared representation learning with distance metric in a unified framework. First, we design a two-pathway deep network pretrained with contrastive loss, and employ double triplet similarity loss for fine-tuning to learn the shared representation for each media type by modeling the relative semantic similarity. Second, the metric network is designed for effectively calculating the cross-media similarity of the shared representation, by modeling the pairwise similar and dissimilar constraints. Compared to the existing methods which mostly ignore the dissimilar constraints and only use sample distance metric as Euclidean distance separately, our UNCSM approach unifies the representation learning and distance metric to preserve the relative similarity as well as embrace more complex similarity functions for further improving the cross-media retrieval accuracy. The experimental results show that our UNCSM approach outperforms 8 state-of-the-art methods on 4 widely-used cross-media datasets.
\end{abstract}

\section{Introduction}
\label{intro}
Recent years, the rapid growth of multimedia data, such as text, image, video and audio, has generated huge requirements for cross-media retrieval. For example, users can submit a text query to retrieve the relevant images or videos that best illustrate the query. The traditional single-media retrieval as \cite{PengCSVT06ClipRetrieval,TypkeISMIR05MusicRetrievalSurvey,YuCSVT08SemanticSubspace} mainly aims to measure the similarity between media instances with the same media type. For examples, some image retrieval methods \cite{DBLP:journals/tois/NieWZC12,DBLP:conf/mm/NieYWHC12} are proposed to predict the search performance and enhance web image reranking by fully exploiting the image content.

\begin{figure}
	\centering
	\includegraphics[width=0.7\textwidth]{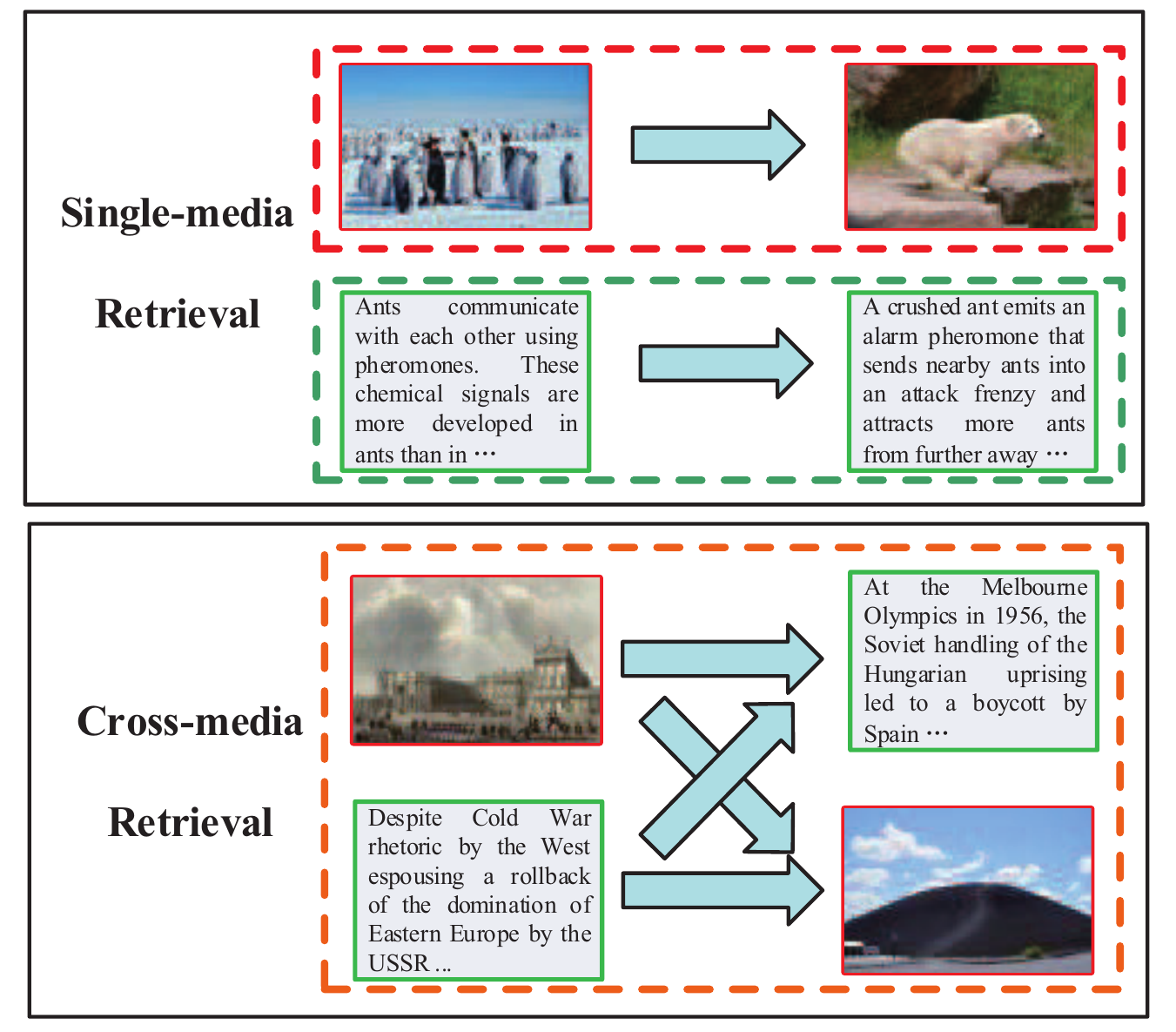}
    \caption{Illustrations of the single-media retrieval and the cross-media retrieval.}
    \label{fig_cross_media}     
\end{figure}

Cross-media retrieval focuses on mining the semantic correlations between data of different media types. The difference between single-media and cross-media retrieval is illustrated in Figure \ref{fig_cross_media}. Comparing with single-media retrieval, cross-media retrieval can provide search results with multiple media types, which makes it more flexible and convenient for users. Under this situation, to fully understand the multimedia data and meet the users' demand for searching whatever they want across different media types, it is increasingly important to model the similarity between different media types for performing cross-media retrieval. However, cross-media retrieval has faced several challenges, including the ``media gap'' between the form of different media features, and ``semantic gap'' between low-level features and high-level semantics. Researchers have proposed different solutions, most of which attempt to present the multimedia data into a common space. They can be mainly divided into two categories which are described as follows. 

One strategy is to adopt traditional linear functions which project the multimedia data into one common space, and obtain the shared representation for each media type \cite{HotelingBiometrika36RelationBetweenTwoVariates,ZhaiAAAI2013JGRHML,ZhaiTCSVT2014JRL}. Then the cross-media similarity can be directly measured by using common distance metric such as Euclidean distance. However, only linear function is used to build the common space, where the complex cross-media correlations cannot be fully captured. Another strategy is inspired by the strong learning ability of deep neural network (DNN). Researchers attempt to use DNN for modeling the correlations between the data of different media types, and so as to get the shared representation. 
However, most of the existing methods \cite{feng12014cross,srivastava2012learning,DBLP:conf/icml/WangALB15} based on DNN only take the pairwise similar constraints and reconstruction information into account. They ignore the dissimilar constraints between different media types, which can provide important hints to cross-media correlation learning. And only sample distance metric is adopted such as Euclidean distance, which limits the accuracy of cross-media retrieval. 


For addressing the above problems, we propose the Unified Network for Cross-media Similarity Metric (UNCSM) to associate the cross-media shared representation learning with distance metric in a unified framework. The main advantages and contributions can be summarized as follows. 
\begin{itemize}
	\item A two-pathway deep network is adopted to model the cross-media semantic correlation, which is pretrained with contrastive loss and fine-tuned with double triplet similarity loss to preserve the relative semantic similarity.
	
	\item A cross-media metric network is designed for effectively calculating cross-media similarity by modeling the pairwise similar and dissimilar constraints. 
\end{itemize}
Our UNCSM approach unifies the representation learning and distance metric to preserve the relative similarity as well as embrace more complex similarity functions for further improving the cross-media retrieval accuracy. The experimental results show that our UNCSM method outperforms 8 state-of-the-art methods on 4 widely-used datasets.

We organize the rest of this paper as follows. We first review the related work on cross-media retrieval in Section 2. Then, our proposed UNCSM approach is presented in Section 3, while the experiments conducted on 4 cross-media datasets and the conclusion of this paper are shown in Section 4 and Section 5.


\section{Related Work}

To measure the cross-media similarity between the data of different media types, most of the existing methods attempt to generate the shared representation for each media types in one common space, which can be divided into two categories: Traditional methods and DNN-based methods.

As for the first strategy, it mainly uses linear function to project the multimedia data into one common space. For example, a straightforward method is to adopt Canonical Correlation Analysis (CCA) \cite{HotelingBiometrika36RelationBetweenTwoVariates}, which is a traditional statistical correlation analysis method, to project the features of different media types into a lower-dimensional common space. Given the training pairs, CCA can find matrices for them, which can make the projected training pairs have maximum correlations with the same dimension to obtain the shared representation. Thus, the simple similarity measurement can be adopted to present cross-media retrieval. Some later works attempt to combine semantic categories to extend CCA, such as \cite{RasiwasiaMM10SemanticCCA}. Besides, Cross-modal Factor Analysis (CFA) \cite{LiMM03CFA} minimizes the Frobenius norm between pairwise data in the transformed domain to learn a common space for different modalities. The joint graph regularized heterogeneous metric learning (JGRHML) \cite{ZhaiAAAI2013JGRHML} is proposed by Zhai et al. to construct the joint graph regularization term using the data in the learned metric space, and this work is further improved as the joint representation learning (JRL) \cite{ZhaiTCSVT2014JRL} by modeling the correlations and semantic information in a unified framework.


In the second strategy, DNN is used to model the correlation between different media types. The strong ability of DNN shown in feature representation learning significantly contributes to the research of single-media retrieval and classification. Some general models such as Stacked Autoencoders (SAE) \cite{SAE}, Deep Belief Network (DBN) \cite{DBLP:journals/neco/HintonOT06} and Deep Boltzmann Machines (DBM) \cite{DBLP:journals/neco/SalakhutdinovH12} are proposed with their learning algorithms for the feature representation learning. Inspired by these, researchers attempt to apply DNN to cross-media retrieval. Bimodal Autoencoders (Bimodal AE) \cite{ngiam32011multimodal} proposed by Ngiam et al. generates the shared representation at the shared code layer taking the speech and visual as input, and it also has a reconstruction layer to reconstruct both two media types. Srivastava and Salakhutdinov propose multimodal DBN \cite{srivastava2012learning} and multimodal DBM \cite{srivastava42012multimodal} to capture the inter-media correlation between different media types at the joint layer. The architectures of Deep CCA \cite{DBLP:conf/icml/AndrewABL13,DBLP:conf/cvpr/YanM15} and Corr-AE \cite{feng12014cross} are similar, consisting of two linked deep encodings. Deep CCA is a non-linear extension of CCA, while Corr-AE can jointly minimize representation learning error and correlation learning error meanwhile with two extensions, namely Corr-Cross-AE and Corr-Full-AE. Besides, CMDN \cite{DBLP:conf/ijcai/PengHQ16} preforms the shared representation learning with multiple deep networks.


The methods mentioned above mainly focus on the cross-media representation learning. As for the cross-media similarity metric, most of the existing methods directly use sample distance metric such as Euclidean distance to measure the similarity of shared representation. Some works such as \cite{ZhaiAAAI2013JGRHML,ZhaiTCSVT2014JRL} attempt to adopt metric learning, but they are also limited in the traditional framework. Inspired by the progress in metric learning based on DNN, we attempt to apply it into cross-media retrieval task. 
However, most of the existing research efforts focus on the single-media scenarios. Wang et al. \cite{DBLP:conf/cvpr/WangSLRWPCW14} propose an efficient triplet sampling algorithm to learn similarity metric for images. It can generate triplet samples to preserve the relative similarity. And each triplet contains three sample with the constraints that the query image needs to be more similar to the positive image than the negative image. Later work as \cite{DBLP:conf/simbad/HofferA15} attempts to learn a better representation by employing triplet network. MatchNet \cite{DBLP:conf/cvpr/HanLJSB15} proposed by Han et al. applies more complex similarity functions than distance metric as Euclidean distance. Although the above methods adopt complex similarity metric, they are all limited by the restricted scenarios and cannot fit for cross-media retrieval. In this paper, we propose a unified framework to associate the cross-media shared representation learning with distance metric, which leads to a better accuracy on cross-media retrieval.


\section{Unified Network for Cross-media Similarity Metric}

\begin{figure}
	\centering
	\includegraphics[width=0.75\textwidth]{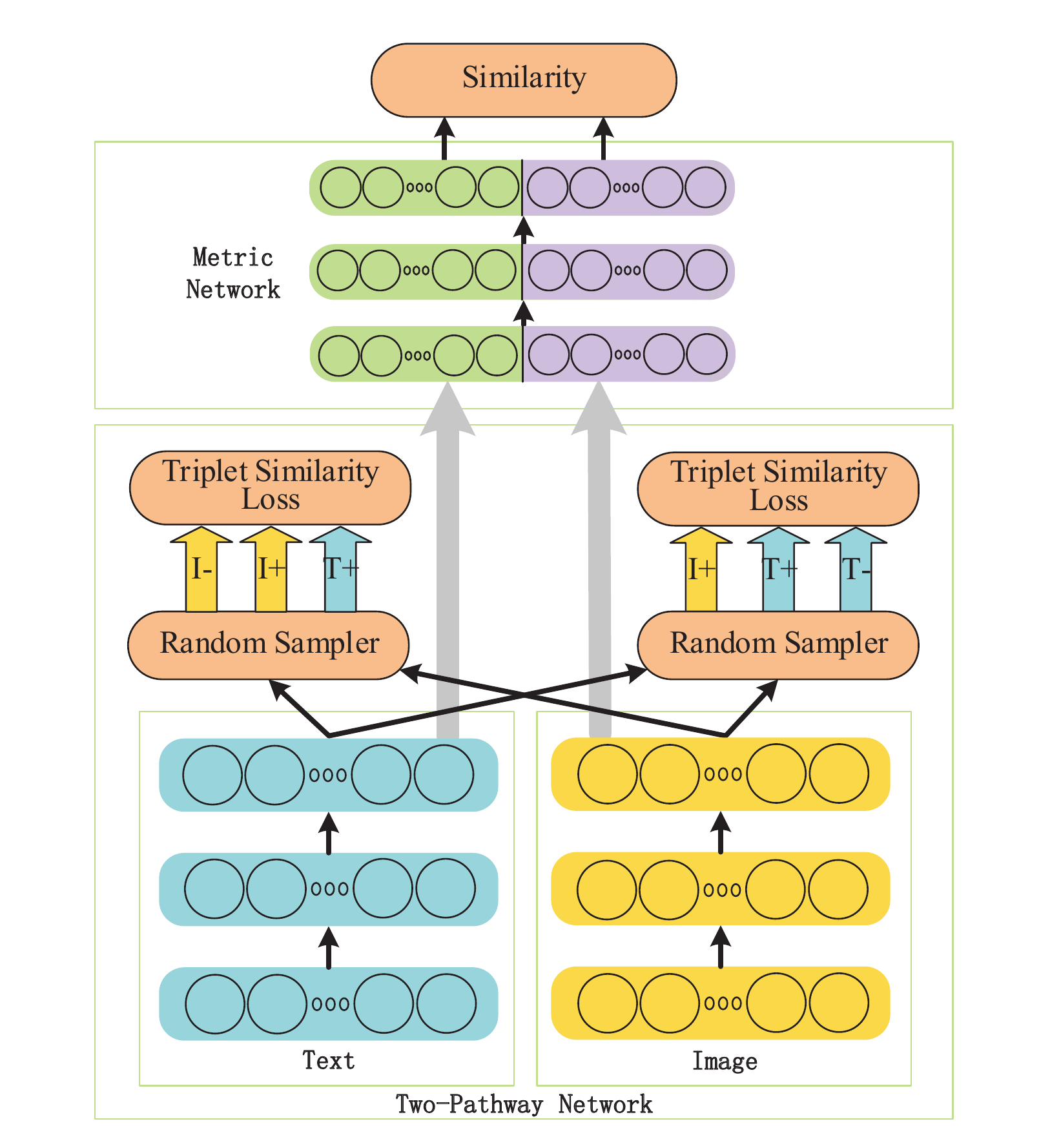}
	\caption{At the bottom is the two-pathway network fine-tuned with double triplet similarity loss embedded on the top. And their output is serve as the input of the metric network directly to calculate the cross-media similarity.}
	\label{fig_network}     
\end{figure}

As shown in Figure \ref{fig_net_pre} and \ref{fig_network}, our UNCSM model consists of two parts as follows. 
\begin{itemize}
	\item A two-pathway deep network to learn the semantic shared representation, which is pretrained with contrastive loss, and two branches of triplet similarity loss embedded on the top of them for the fine-tune stagee.
	
	\item A metric network for modeling the pairwise similar and dissimilar constraints to calculate the cross-media similarity. 
	
\end{itemize}
Formally, $D^{(i)}=\left \{x_{p}^{(i)},y_{p}^{(i)}\right \}_{p=1}^{m}$ denotes the image data which has $m$ image instances. $x_{p}^{(i)}\in \mathbb{R}^{d^{(i)}}$ is the $p$-th image data with the dimensional number $d^{(i)}$ and the corresponding label $y_{p}^{(i)}$. The definition of text data $D^{(t)}=\left \{x_{p}^{(t)},y_{p}^{(t)}\right \}_{p=1}^{n}$ is similar to image data.

\subsection{Two-pathway Network}


\begin{figure}
	\centering
	\includegraphics[width=0.75\textwidth]{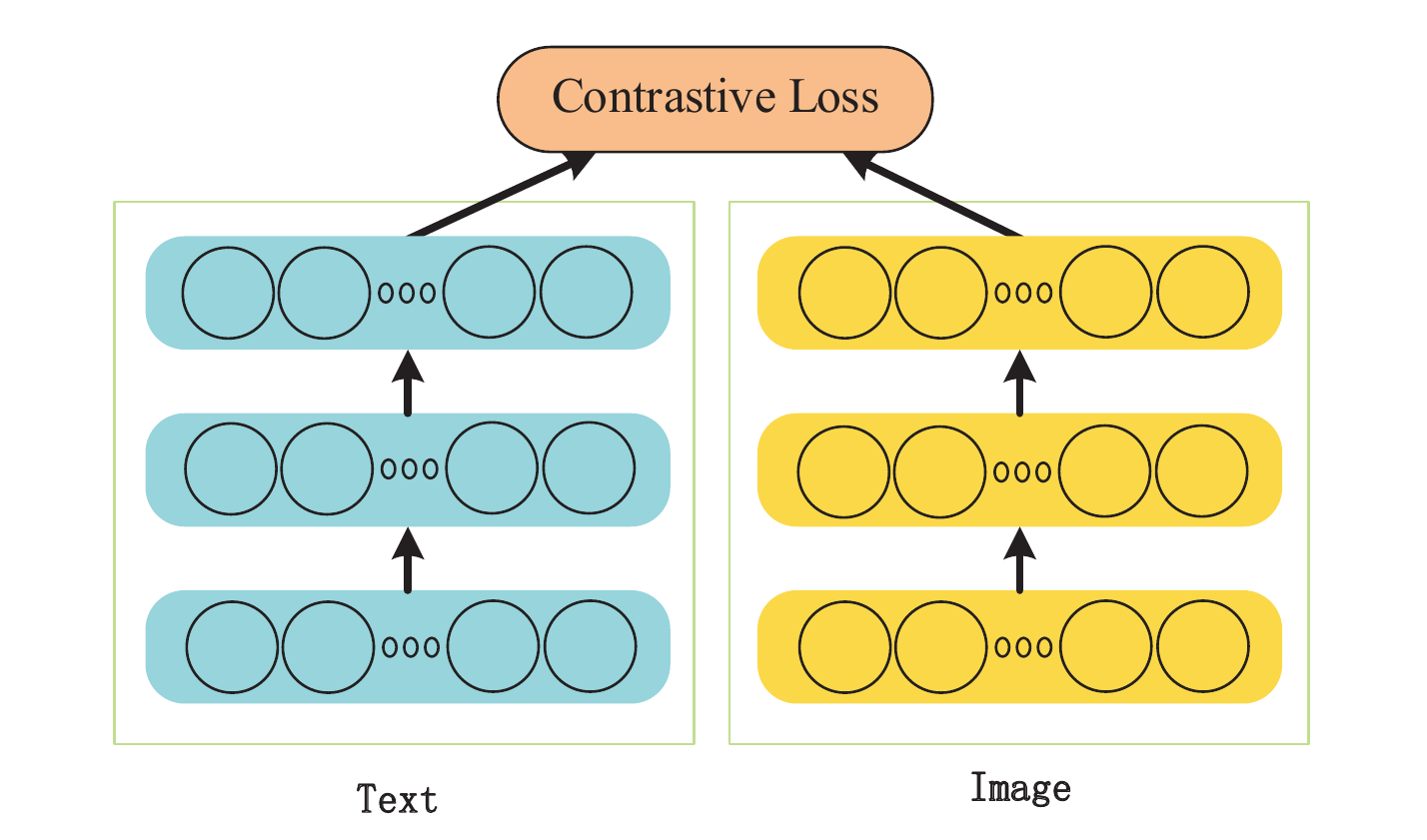}
	\caption{Pretrain network with contrastive loss.}
	\label{fig_net_pre}     
\end{figure}

\textit{\textbf{Pretraining:}} Deep neural network has shown strong ability in modeling the correlations between different media types, but it may easily fall into local minimum. Therefore, to find a good region in parameter space, we first adopt pretraining in shared representation learning.

For modeling the pairwise similar and dissimilar correlations between different media types, the contrastive loss is adopted to pretrain the two-pathway network with the following consideration, which is shown in Figure \ref{fig_net_pre}. The image and text instances with the same class label should be closer in distance, while there should be larger distance between the image and text instances which have different class labels. So we define the contrastive loss as follows:
 \begin{equation}
 \label{equ_con}
 C(p,q)=
 \begin{cases}
 \left \| f_{i}(x_{p}^{(i)})-f_{t}(x_{q}^{(t)}) \right \|^{2}  &  y_{p}^{(i)}=y_{q}^{(t)} \\
 max(0,\lambda-\left \| f_{i}(x_{p}^{(i)})-f_{t}(x_{q}^{(t)}) \right \|)^{2} & y_{p}^{(i)}\neq y_{q}^{(t)}
 \end{cases},
 \end{equation}
where $f_{i}(.)$ and $f_{t}(.)$ denote the non-linear mapping through each pathway of the network. And the margin parameter is set to be $\lambda$. 

\textit{\textbf{Fine-tuning:}} After pretrain stage, the parameters fall into a good region in parameter space. However, they still need to be optimized. So we adopt fine-tune process to refine the parameters to the local optimum. Inspired by the idea that the triplet network \cite{DBLP:conf/cvpr/WangSLRWPCW14} can preserve the relative similar and dissimilar constraints by generating the triplet samples, we employ the double triplet similarity loss for each media type at fine-tune stage.


Specially, we propose an online triple sampling strategy to generate triplet samples for each media type using the feature extracted from the separate two-pathway networks. For image data $f_{i}(x_{p}^{(i)})$ extracted from image pathway, the triplet samples are organized as $(I^{+},T^{+},T^{-})$ considering the relative similar and dissimilar constraints that the image sample $I^+$ is similar to positive text sample $T^+$ and dissimilar to the negative text sample $T^-$. And the organization of triplet samples for text data $f_{t}(x_{p}^{(t)})$ is similar as $(T^{+},I^{+},I^{-})$ satisfying the constraints that the text sample $T^+$ has its similar image sample $I^+$ and its dissimilar image sample $I^-$. These relative similar and dissimilar constraints are measured by their corresponding label $y_{p}^{(i)}$ and $y_{p}^{(t)}$.
So the double triplet similarity loss are defined as follows:
\begin{gather}
S_{i}(I^{+},T^{+},T^{-})=max(0, \left \| f_{i}(x^{I+})-f_{t}(x^{T+}) \right \|_{2}^{2} 
-\left \| f_{i}(x^{I+})-f_{t}(x^{T-}) \right \|_{2}^{2}+\alpha), \label{equ:tri_img1}\\
\label{equ:tri_img2}
S_{t}(T^{+},I^{+},I^{-})=max(0, \left \| f_{t}(x^{T+})-f_{i}(x^{I+}) \right \|_{2}^{2} 
-\left \| f_{t}(x^{T+})-f_{i}(x^{I-}) \right \|_{2}^{2}+\beta),
\end{gather}
where $\left \| . \right \|_{2}$ is the $\mathcal{L}_{2}$ norm and the margin parameters are set as $\alpha$ and $\beta$ here, which makes the difference between the distance of similar pairs and dissimilar pairs is larger than the defined margin.

It should be noted that the double triplet similarity loss is treated as two loss branches embedded on the top of the two-pathway network, which has a fully-connected layer using sigmoid non-linearity for each branch. So the gradients from each loss branch are summed together on the top of each pathway network for updating the parameters by back propagation.

Using the mapping function $f_{i}(.)$ and $f_{t}(.)$, we can get the optimized semantic shared representation $f_{i}(X^{(i)})$ and $f_{t}(X^{(t)})$ which preserve the relative similarity between different media types, and serve as the input of the cross-media metric network denoted as $M^{(i)}$ and $M^{(t)}$.

\subsection{Cross-media Metric Network}

It is not sufficient enough to adopt the simple distance metric as Euclidean distance for calculating the complex similarity of cross-media data. Inspired by \cite{DBLP:conf/cvpr/HanLJSB15}, we design a network of cross-media distance metric for further calculating the cross-media similarity, which can model the pairwise similar and dissimilar constraints. The metric network can learn the similarity functions from the positive and negative sample pairs, and finally generate the cross-media similarity.

In the UNCSM, the metric network is a feed-forward network which consists of three fully-connected layers using Rectified Linear Units (ReLU) as non-linearity, and there is a Softmax layer on the top of the network. 
The metric network takes the concatenation of a pair of image and text features as input, and the output includes two values in $[0,1]$ from the two units in the Softmax layer, which can be used as the network's estimate of the similarity. 
Specifically, the image and text feature, used to concatenate each pairwise sample $q_{i}$, are selected from the shared representation $M^{(i)}$ for image media and $M^{(t)}$ for text media obtained by the two-pathway network, using a random sampler. These randomly selected pairwise samples can be divided into the positive or negative pairs according to their class labels.
The generated training set consists of $n$ pairs of features and we minimize the cross-entropy error as follows:
\begin{gather}
E=-\frac{1}{n}\sum_{i=1}^{n}(p_{i}log(\hat{p_{i}})+(1-p_{i})log(1-\hat{p_{i}})),
\end{gather}
where $p_{i}$ is denoted as the 0/1 label for the input pair $q_{i}$ and $p_{i}=1$ if the input sample pair belongs to the same class. The Softmax activations $\hat{p_{i}}$ and $1-\hat{p_{i}}$ are computed on the values of the two nodes $h_{0}(q_{i})$ and $h_{1}(q_{i})$ in the Softmax layer as follows:
\begin{gather}
\hat{p_{i}}=\frac{e^{h_{1}(q_{i})}}{e^{h_{0}(q_{i})}+e^{h_{1}(q_{i})}}.
\end{gather}

Finally, we use calculated $\hat{p_{i}}$ to estimate whether the input pair belongs to the same class or not, which is further used as the final learned cross-media similarity metric, and the cross-media retrieval can be performed by ranking the learned similarity.
It should be noted that our training process is not an end-to-end training style, which is divided into three separate stages to ensure that each part of our model has good performance. While in testing stage, it is actually a unified process, which takes the features of different media types as input and directly obtains the cross-media similarity as output.

\section{Experiment}

We conduct experiments on 4 widely-used cross-media datasets. And 8 existing cross-media methods are compared on two-retrieval tasks to verify the effectiveness of our approach. First, 4 datasets used in the experiments will be introduced, which are Wikipedia, NUS-WIDE, NUS-WIDE-10k and Pascal Sentences, along with the feature extraction as well as the dataset partition. The descriptions for compared methods and evaluation metric will be shown next. And then we will give the network details and parameter analysis. Finally, the experimental results and analysis is given. 


\begin{figure}
	\centering
	\includegraphics[width=0.75\textwidth]{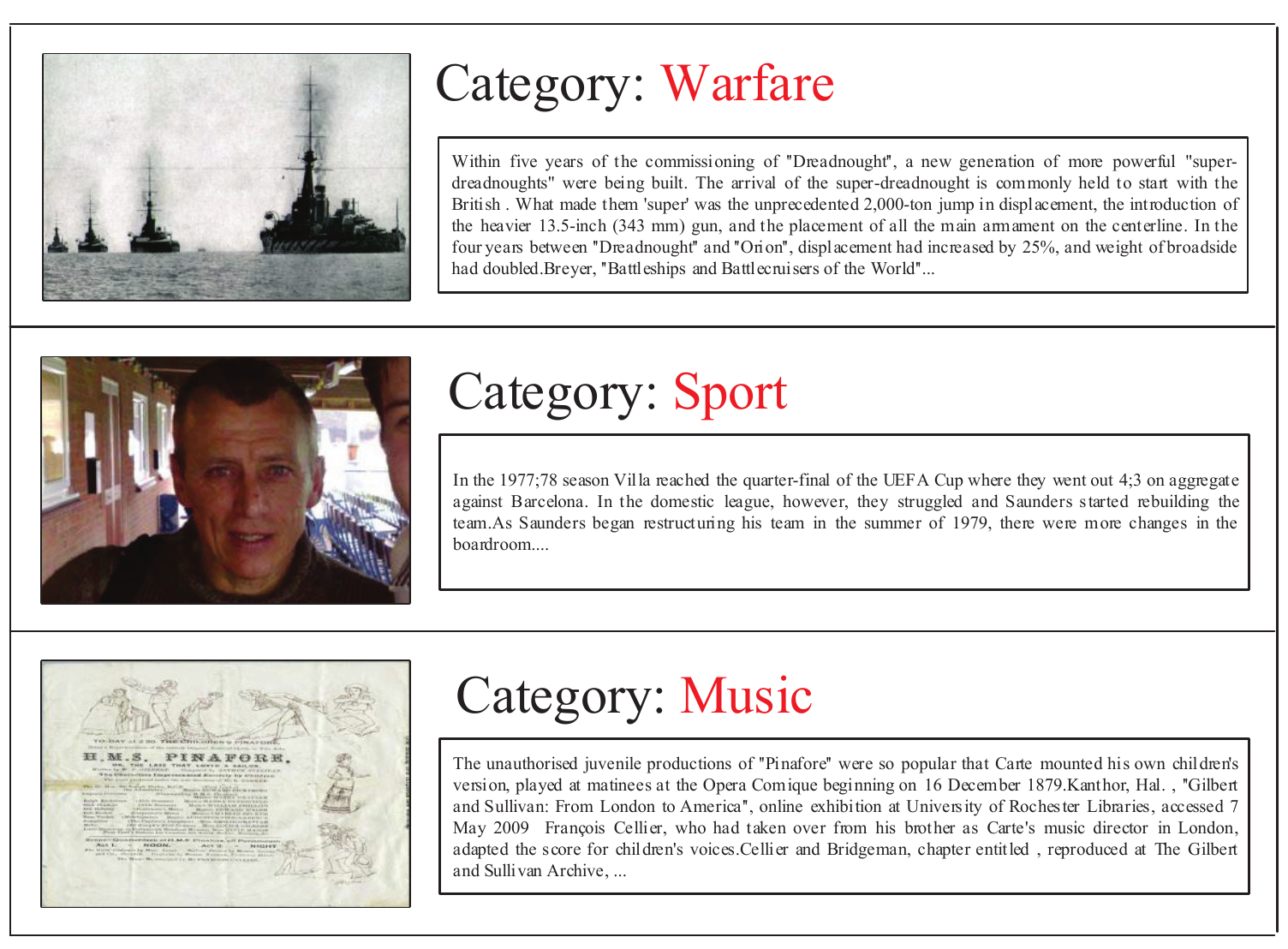}
	\caption{Examples of 3 categories from the Wikipedia dataset.}
	\label{fig_wiki_ex}     
\end{figure}

\subsection{Datasets and Feature Extraction}

Totally 4 datasets are selected for the experiments, which are widely-used in cross-media retrieval. The brief description of them as well as the feature extraction strategy will be given as follows. 

{\textbf {Wikipedia dataset}} \cite{RasiwasiaMM10SemanticCCA} is selected from the Wikipedia's ``feature articles'' with 10 classes, which contains totally 2,866 documents in form of image/text pair. Some examples of 3 categories selected from Wikipedia dataset are shown in Figure \ref{fig_wiki_ex}. We randomly split the dataset into three part following \cite{feng12014cross}. 2,173 documents are randomly selected as training set. Testing set contains 462 documents and 231 documents are in validation set. The image feature has three parts: 1,000 dimensional Pyramid Histogram of Words (PHOW) \cite{PHOW2007}, 512 dimensional GIST \cite{GIST2001}, and 784 dimensional MPEG-7 \cite {MPEG72001}, which concatenates into 2,296 dimensional feature. And the text feature uses 3,000 dimensional bag of words vector.

{\textbf {NUS-WIDE dataset}} \cite{NUSWIDE} is a web image dataset which is created by NUS's Lab for media search, containing images and their corresponding text tags crawled from Flickr through its public API. There are 269,648 images in NUS-WIDE dataset. Because the classes have overlaps, we select 10 largest classes from it including about 100K image/text pairs with unique class label. The image feature has 1,134 dimensions with several parts as follows: 64 dimensional color histogram, 144 dimensional color correlogram, 73 dimensional edge direction histogram, 128 dimensional wavelet texture, 225 dimensional block-wise color moments and 500 dimensional SIFT-based bag of words features. The text feature is represented as 1,000 dimensional bag of words vector. The dataset is also randomly split into three parts. There are 58,620 documents in training set, 33,955 documents in testing set and 5,000 documents in validation set.

{\textbf {NUS-WIDE-10k dataset}} \cite{NUSWIDE} is a subset of NUS-WIDE dataset, which totally has 10,000 image/text pairs selected from the 10 largest classes in NUS-WIDE dataset, and 1,000 image/text pairs are contained in each class. The image and text features are the same with NUS-WIDE dataset. And following \cite{feng12014cross}, 8,000 documents are randomly selected as training set, testing set contains 1,000 documents and 1,000 documents are in validation set.

{\textbf {Pascal Sentences}} \cite{Pascal} is selected from 2008 PASCAL development kit, which contains 1,000 image/text pairs organized into 10 categories. And each image instance has 5 sentences as description. There are 800 documents selected as training set, 100 documents for testing set and 100 documents for validation set, also following \cite{feng12014cross}. We extract the same image feature with Wikipedia dataset, and 1,000 dimensional bag of words vector as text representation.

\subsection{Compared Methods}

We compare our UNCSM approach with 8 existing cross-media methods. CCA, CFA and KCCA are the classical baselines, while the others are all based on DNN. The source codes of Bimodal AE, Multimodal DBN and Corr-AE are available from \cite{feng12014cross}, and the source codes of DCCA and DCCAE are from \cite{DBLP:conf/icml/WangALB15}. We will introduce these 8 compared methods briefly as follows:

\begin{itemize}
	\item {\bf CCA} \cite{HotelingBiometrika36RelationBetweenTwoVariates}. CCA projects the data with two media types into a common subspace by maximizing the pairwise correlations.
	
	\item {\bf CFA} \cite{LiMM03CFA}. CFA learns a common space for different modalities by minimizing the Frobenius norm between pairwise data in the transformed domain.
	
	\item {\bf KCCA} \cite{DBLP:journals/neco/HardoonSS04}. KCCA uses kernel functions to project the data of different media types into high-dimensional feature space. 
	Polynomial kernel (Poly) and radial basis function (RBF) are used as the kernel functions in our experiments.
	
	\item {\bf DCCA} \cite{DBLP:conf/icml/AndrewABL13}. DCCA is a non-linear extension of CCA, which has two separate subnetworks and maximizes the correlation between the output layers.
	
	\item {\bf DCCAE} \cite{DBLP:conf/icml/WangALB15}. DCCAE consists of two autoencoders, which is optimized by the combination of canonical correlation and reconstruction errors.
	
	\item {\bf Bimodal AE} \cite{ngiam32011multimodal}. Bimodal AE can generate the shared representation at a shared code layer, and it is also required to reconstruct both two media types at the reconstruction layer.
	
	\item {\bf Multimodal DBN} \cite{srivastava2012learning}. Multimodal DBN adopts two separate DBN to model the distribution over multiple media types and learns the joint representation by a joint RBM on the top of them. 
	
	\item {\bf Corr-AE} \cite{feng12014cross}. Corr-AE uses two linked subnetworks to model the correlation loss as well as reconstruction error, which has two extensions, namely Corr-Cross-AE and Corr-Full-AE. In our experiments, we compare the highest result of three models.
	
\end{itemize}

\subsection{Evaluation Metric}

To evaluate the performance of our UNCSM approach, two cross-media retrieval tasks are conducted on the above 4 datasets, which are image retrieval by a text query (Text$\rightarrow$Image) and text retrieval by an image query (Image$\rightarrow$Text). We evaluate the experiment results with the mean average precision (MAP) and PR (precision-recall) curves. MAP is widely-used in information retrieval literature. And the average precision (AP) is defined as follows:
\begin{equation}
AP = \frac 1 M \sum_{k=1}^n P_k \times Rel_k,
\end{equation}
where $M$ is the total relevant item number in testing set with the size of $n$. Here $P_k$ is the precision at $k$-th position, while $Rel_k$ is 1 when the item on the corresponding position is relevant and 0 otherwise. In our experiments, we compute the MAP score of \textit{all the returned results} for both the compared methods and our method, instead of MAP score only calculated for the top 50 results in \cite{feng12014cross}. The PR curve indicates the precisions at certain level of recall, while we also calculate the PR curves of all returned results.

\begin{figure}
	\centering
	\includegraphics[width=0.6\textwidth]{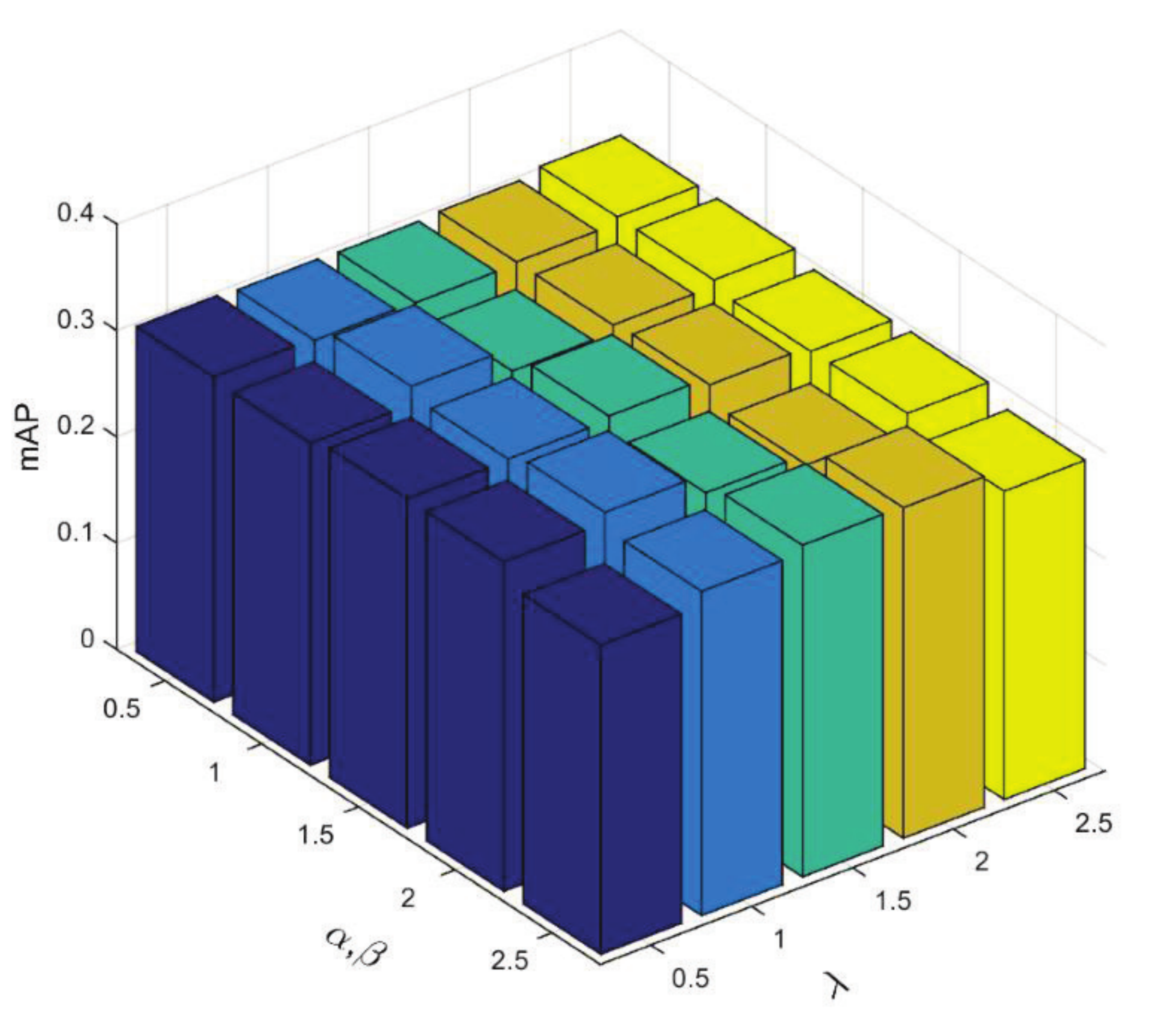}
	\caption{MAP score with respect to parameter variations on Wikipedia dataset.}
	\label{fig_margin}     
\end{figure}

\subsection{Network Details and Parameter Analysis}

Generally, the features of different media types are presented in different forms and the statistical properties of them are also different to each other, which is hard to directly adopt our UNCSM model to learn the cross-media similarity metric between different media types. Therefore, we adopt a preprocessing phase, which first employs two DBN to model the distribution over the features of different media types separately, and each DBN has two layers. For the image features, we use the Gaussian Restricted Boltzmann Machine (RBM) \cite{DBLP:conf/nips/WellingRH04}. And Replicated Softmax model \cite{DBLP:conf/nips/SalakhutdinovH09} is used for the text features. The first layer of DBN for image input features has 2048 hidden units and the second layer has 1024 hidden units. As for text input features, there are 1024 hidden units on both two layer of DBN. We also use two feed-forward network using Softmax loss to further optimize the features for each media type, which can preserve the intra-media information with each media type. 

As for the two-pathway network illustrated in Figure \ref{fig_net_pre} and \ref{fig_network} in both pretrain and fine-tune stages, the first layer on each pathway has the dimensional number of 1024. The dimensions of the second layer reduce to half of the first layer having 512 dimensions, and the third layer has 256 dimensions which is also half of the former layer. The fully-connected layer on double triplet similarity loss branch also has the dimensional number of 256. It should be noted that there are three major margin parameters involved in the formulations of this part, which are $\lambda$, $\alpha$ and $\beta$ in eqn \ref{equ_con}, \ref{equ:tri_img1} and \ref{equ:tri_img2}. We construct a baseline experiment to show the impact of parameters. Specifically, $\lambda$ is the margin parameter for the distance metric of dissimilar pair, while $\alpha$ and $\beta$ are the margin parameters for the difference between the distance of similar pairs and dissimilar pairs. We set them with different values ranging from 0.5 to 2.5, where $\alpha$ and $\beta$ are set equally here for the equal treatment of image and text in two triplet loss. And the average MAP scores with cosine distance metric are computed on Wikipedia dataset. The results are presented in Figure \ref{fig_margin}, which shows the experiment results are less sensitive to these parameters. Thus, the three margin parameters are all set to be 1 in the following experiments.

\begin{table}
	\centering
	\caption{The MAP scores on Wikipedia dataset.}
	\label{table:res_wiki}       
	\begin{tabular}{llll}
		\hline\noalign{\smallskip}
		Method & Image$\rightarrow$Text & Text$\rightarrow$Image & Average  \\
		\noalign{\smallskip}\hline\noalign{\smallskip}
		CCA & 0.124 & 0.120& 0.122 \\
		CFA & 0.236 & 0.211& 0.224 \\
		KCCA(Poly) & 0.200 & 0.185& 0.193\\
		KCCA(RBF) & 0.245 & 0.219& 0.232 \\
		Bimodal AE & 0.236 & 0.208& 0.222 \\
		Multimodal DBN & 0.149 & 0.150& 0.150 \\
		DCCA & 0.248 & 0.221& 0.235 \\
		Corr-AE & 0.280 & 0.242& 0.261 \\
		DCCAE & 0.260 & 0.233& 0.247 \\
		\textbf{UNCSM} & \textbf{0.360} & \textbf{0.310} & \textbf{0.335} \\
		\noalign{\smallskip}\hline
	\end{tabular}
\end{table}

\begin{table}
	\centering
	\caption{The MAP scores on NUS-WDIE Dataset.}
	\label{table:res_nusL}       
	\begin{tabular}{llll}
		\hline\noalign{\smallskip}
		Method & Image$\rightarrow$Text & Text$\rightarrow$Image & Average  \\
		\noalign{\smallskip}\hline\noalign{\smallskip}
		CCA & 0.242 & 0.240& 0.241 \\
		CFA & 0.256 & 0.288& 0.272 \\
		KCCA(Poly) & 0.242 & 0.231& 0.237\\
		KCCA(RBF) & 0.258 & 0.275& 0.267 \\
		Bimodal AE & 0.266 & 0.300& 0.283 \\
		Multimodal DBN & 0.238 & 0.261& 0.250 \\
		DCCA & 0.367 & 0.363& 0.365 \\
		Corr-AE & 0.323 & 0.354& 0.339 \\
		DCCAE & 0.369 & 0.362& 0.366 \\
		\textbf{UNCSM} & \textbf{0.514} & \textbf{0.477} & \textbf{0.496} \\
		\noalign{\smallskip}\hline
	\end{tabular}
\end{table}

In the metric network, since the concatenation of feature obtained from the bottom two-pathway network has 512 dimensions, the dimensional number of the three fully-connected layers is also 512 and the Softmax layer has 2 dimensions for the final similarity metric. We adopt random sampling strategy to generate the similar and dissimilar pairs with the ratio of 1:1 with about 300K samples. It should be noted that all the above parameters and number of samples are fit for the Wikipedia dataset, while the parameters should be adjusted to suit other dataset. 
Besides, for the fact that the number of parameters in the deep model is associated with the setting of network architecture and the network architecture of our model as mentioned above is not complex, there are about 5K parameters totally for our model to learn and this number is relatively small, which can be trained successfully with small amount of training data. Moreover, we train the whole network using a base learning rate 0.001 by stochastic gradient descent, and the weight decay parameter is 0.004. Also the training process can converge fast with less than 10K iterations for each part, and does not cost much training time which is within 1 hour totally.

%

\begin{table}
	\centering
	\caption{The MAP scores on NUS-WDIE-10k Dataset.}
	\label{table:res_nus}       
	\begin{tabular}{llll}
		\hline\noalign{\smallskip}
		Method & Image$\rightarrow$Text & Text$\rightarrow$Image & Average  \\
		\noalign{\smallskip}\hline\noalign{\smallskip}
		CCA & 0.120 & 0.120& 0.120 \\
		CFA & 0.211 & 0.188& 0.200 \\
		KCCA(Poly) & 0.150 & 0.149& 0.150\\
		KCCA(RBF) & 0.232 & 0.213& 0.223 \\
		Bimodal AE & 0.159 & 0.172& 0.166 \\
		Multimodal DBN & 0.158 & 0.130& 0.144 \\
		DCCA & 0.219 & 0.210& 0.215 \\
		Corr-AE & 0.223 & 0.227& 0.225 \\
		DCCAE & 0.227 & 0.216& 0.222 \\
		\textbf{UNCSM} & \textbf{0.312} & \textbf{0.354} & \textbf{0.333} \\
		\noalign{\smallskip}\hline
	\end{tabular}
\end{table}

\begin{table}
	\centering
	\caption{The MAP scores on Pascal Sentences Dataset.}
	\label{table:res_pascal}       
	\begin{tabular}{llll}
		\hline\noalign{\smallskip}
		Method & Image$\rightarrow$Text & Text$\rightarrow$Image & Average  \\
		\noalign{\smallskip}\hline\noalign{\smallskip}
		CCA & 0.099 & 0.097& 0.098 \\
		CFA & 0.187 & 0.216& 0.202 \\
		KCCA(Poly) & 0.207 & 0.191& 0.199\\
		KCCA(RBF) & 0.233 & 0.249& 0.241 \\
		Bimodal AE & 0.245 & 0.256& 0.251 \\
		Multimodal DBN & 0.197 & 0.183& 0.190 \\
		DCCA & 0.252 & 0.247& 0.250 \\
		Corr-AE & 0.268 & 0.273& 0.271 \\
		DCCAE & 0.253 & 0.241& 0.247 \\
		\textbf{UNCSM} & \textbf{0.304} & \textbf{0.282} & \textbf{0.293} \\
		\noalign{\smallskip}\hline
	\end{tabular}
\end{table}

\subsection{Experimental Results}

Now we will give the experimental results and analysis for both the compared methods and our UNCSM approach. The MAP scores on Wikipedia, NUS-WIDE, NUS-WIDE-10k and Pascal Sentences dataset are shown in Tables \ref{table:res_wiki} to \ref{table:res_pascal}.

\begin{figure}
	\centering
	\includegraphics[width=0.72\textwidth]{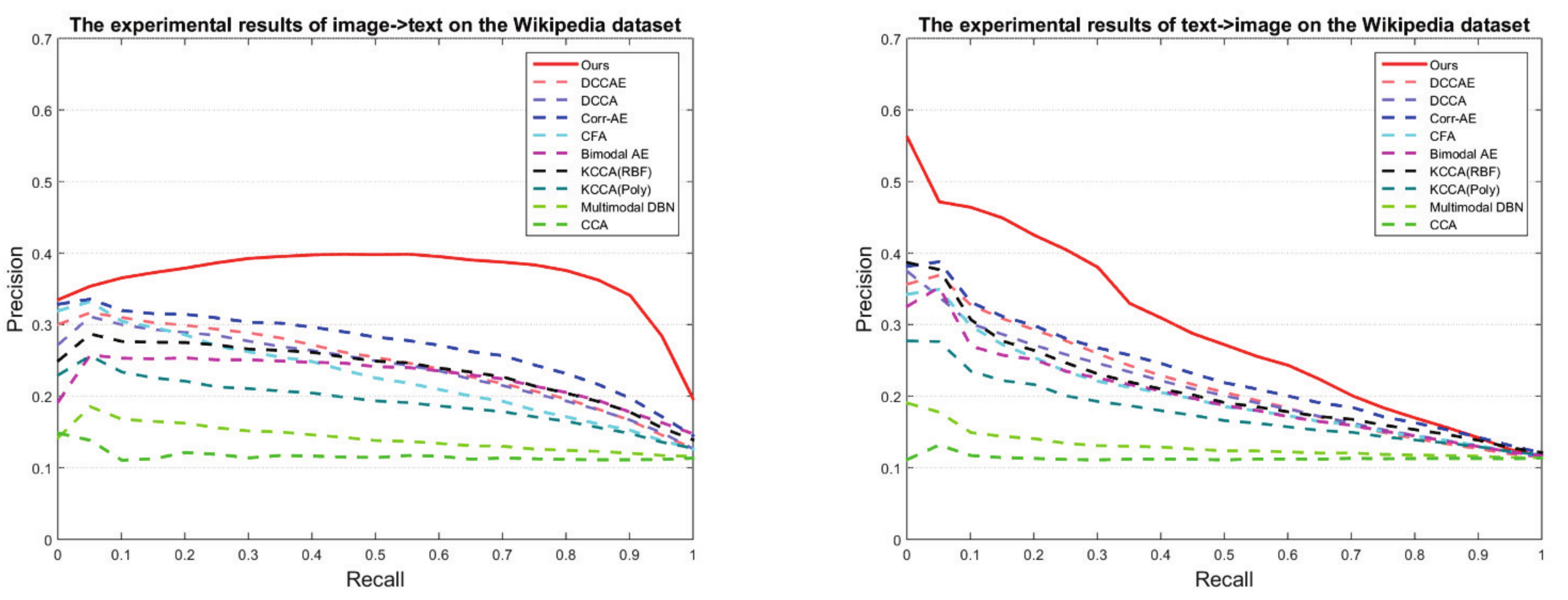}
	\caption{The PR curves on Wikipedia dataset.}
	\label{fig_pr_wiki}     
\end{figure}

\begin{figure}
	\centering
	\includegraphics[width=0.72\textwidth]{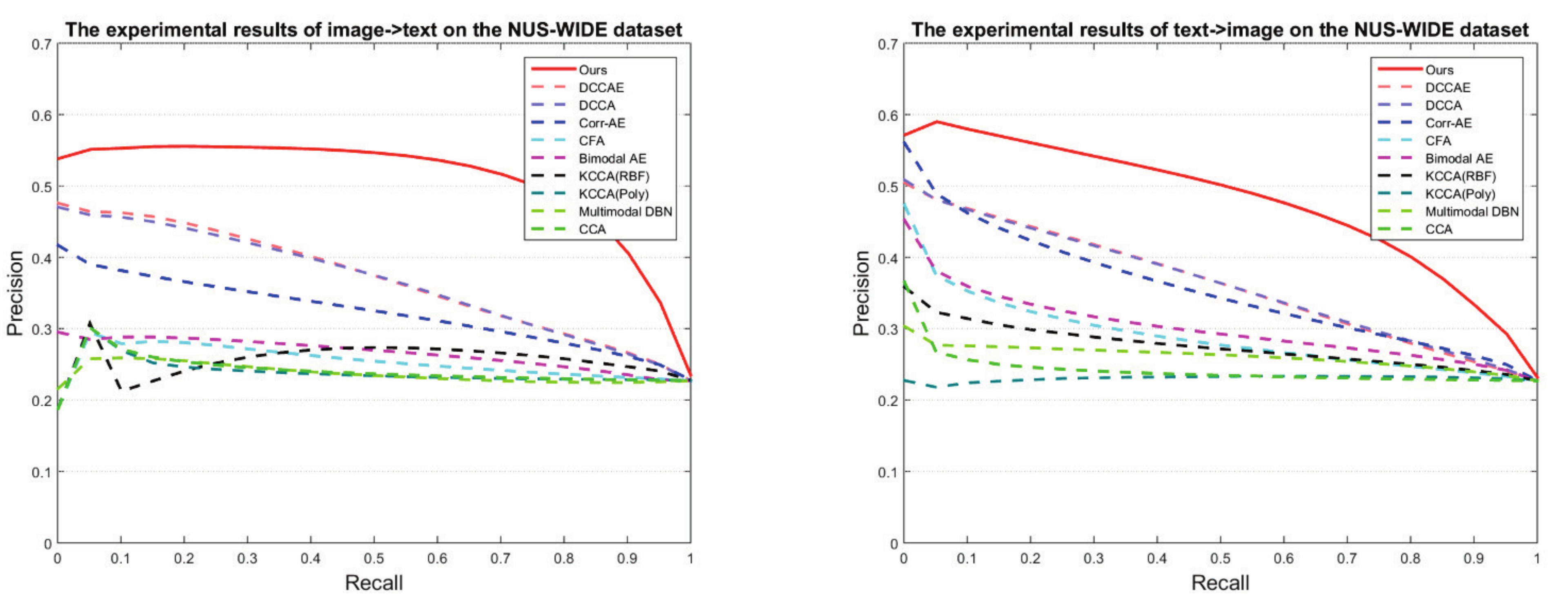}
	\caption{The PR curves on NUS-WIDE dataset.}
	\label{fig_pr_nusL}     
\end{figure}

\begin{figure}
	\centering
	\includegraphics[width=0.72\textwidth]{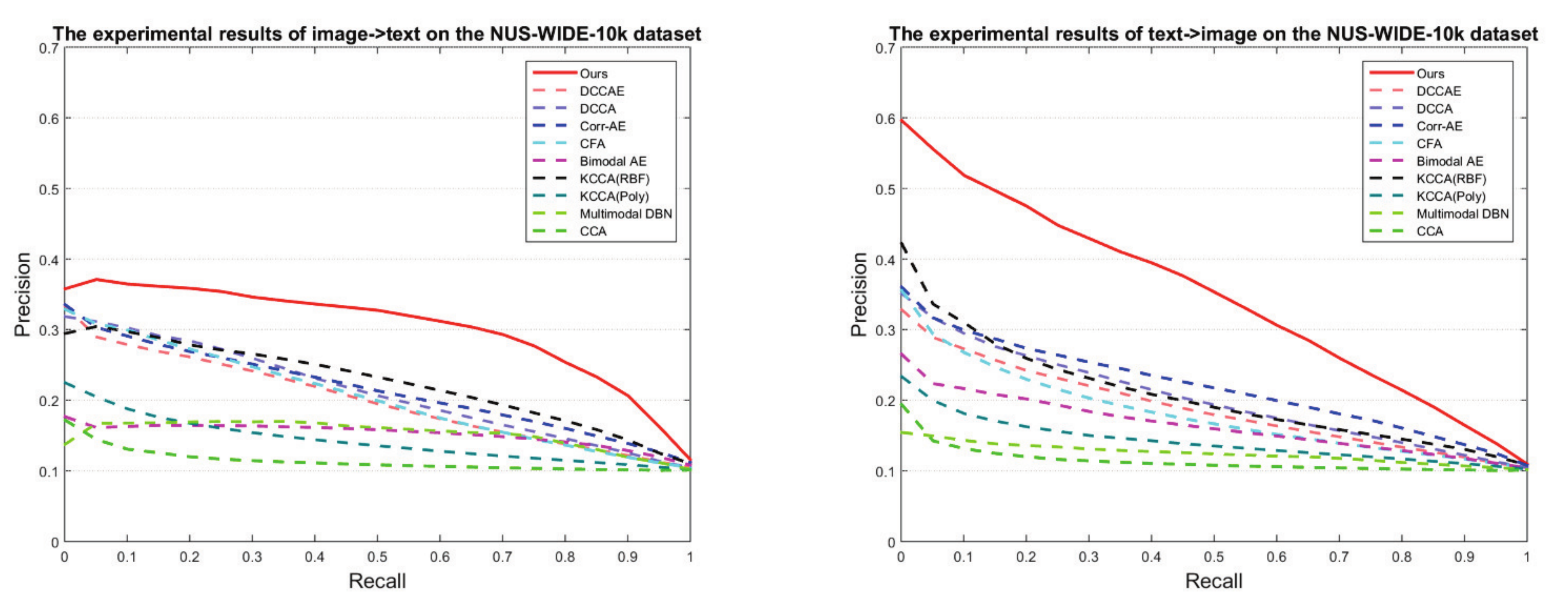}
	\caption{The PR curves on NUS-WIDE-10k dataset.}
	\label{fig_pr_nus}     
\end{figure}

\begin{figure}
	\centering
	\includegraphics[width=0.72\textwidth]{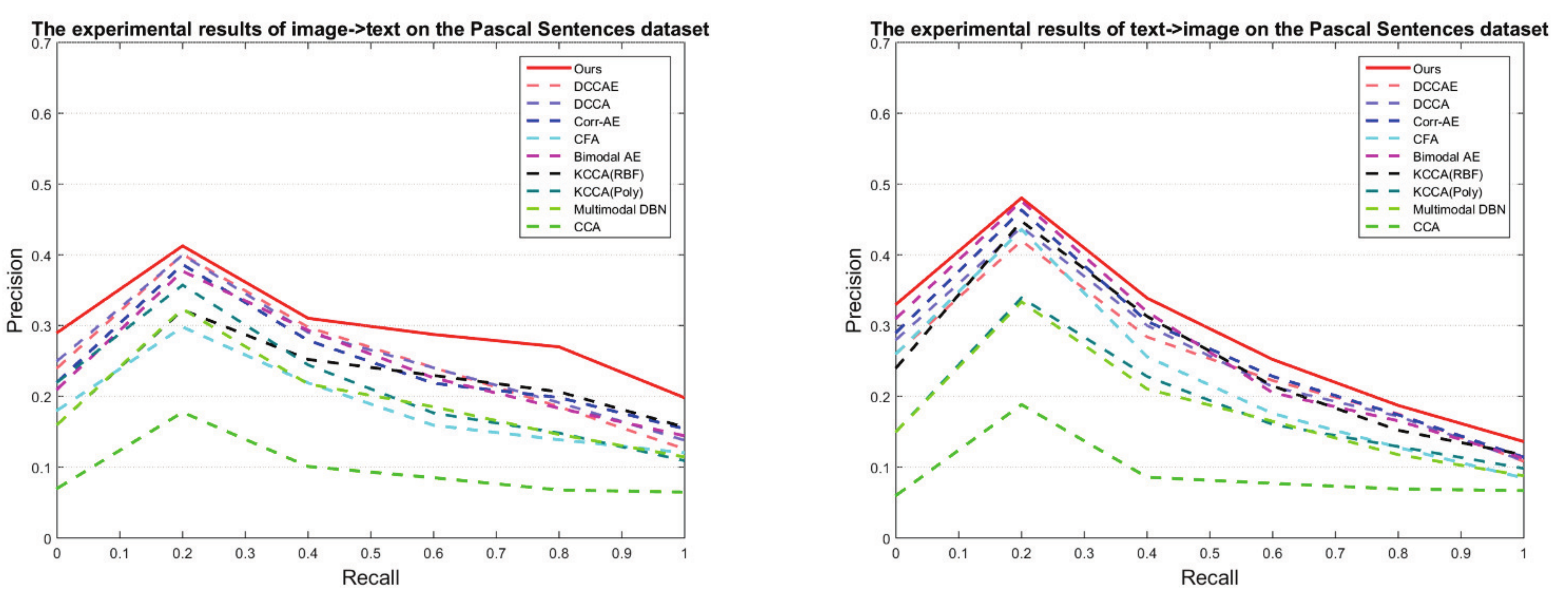}
	\caption{The PR curves on Pascal Sentences dataset.}
	\label{fig_pr_pascal}     
\end{figure}

\begin{figure}
	\centering
	\includegraphics[width=0.95\textwidth]{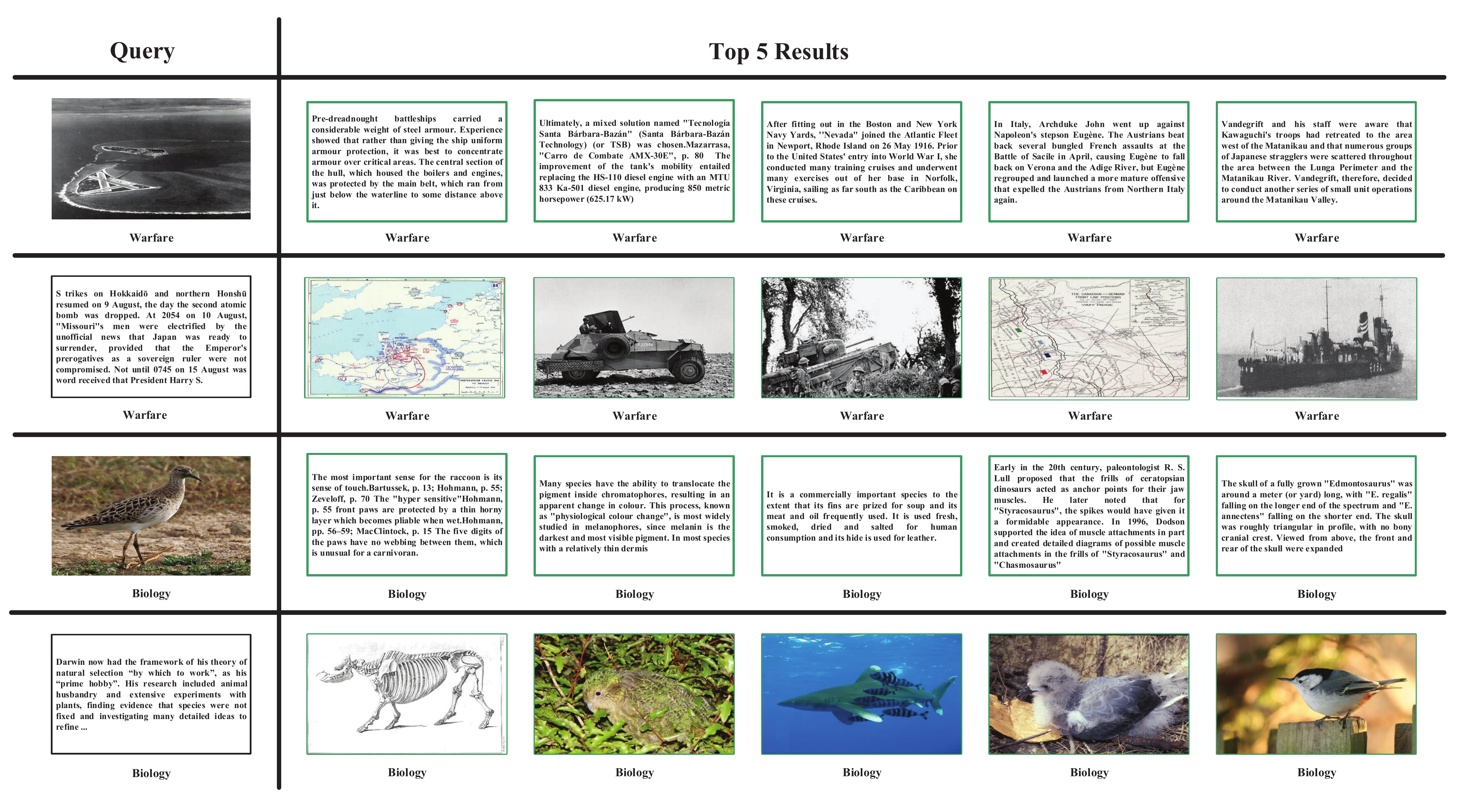}
	\caption{Some retrieval results on Wikipedia dataset. There are four example queries on the left row, and top 5 results are shown on the right. It should be noted that all these results shown above are correct.}
	\label{fig_wiki_res}     
\end{figure}

Table \ref{table:res_wiki} shows the MAP scores of the two retrieval tasks and their average results on Wikipedia dataset. Our UNCSM approach ourperforms all the compared methods on two tasks, achieving significant improvement from 0.261 to 0.335 on average result. The performance of DNN-based methods is not stable that Multimodal DBN is outperformed by most of the compared methods except CCA, while Corr-AE has the best result among the compared methods. And some examples of retrieval results on Wikipedia dataset are shown in Figure \ref{fig_wiki_res}.
Besides, the results of two retrieval tasks on other three datasets in Tables \ref{table:res_nusL} to \ref{table:res_pascal} show similar trends as in Wikipedia dataset, and our UNCSM approach still keeps the best.
It should be noted that the MAP scores on NUS-WIDE dataset is higher than NUS-WIDE-10k dataset, which indicates that our model is more suitable for large-scale dataset and can achieve better accuracy with more training data.
The PR curves on the 4 datasets are shown in Figures \ref{fig_pr_wiki} to \ref{fig_pr_pascal}, and our proposed approach shows clear advantage over all the compared methods, which demonstrates its effectiveness. 

\begin{table}
	\centering
	\caption{MAP scores of Cosine distance metric and our Metric Network on the shared representation obtained from two-pathway network.}
	\label{table:res_metric}       
	\begin{tabular}{llll}
		\hline\noalign{\smallskip}
		Dataset & Task & Cosine Distance & Metric Network  \\
		\noalign{\smallskip}\hline\noalign{\smallskip}
		& Image$\rightarrow$Text & 0.301 & \textbf{0.360} \\
		Wikipedia & Text$\rightarrow$Image & 0.290 & \textbf{0.310} \\
		& Average & 0.296 & \textbf{0.335}\\
		\hline
		& Image$\rightarrow$Text & 0.470 & \textbf{0.514}\\
		NUS-WIDE & Text$\rightarrow$Image & 0.455 & \textbf{0.477}\\
		& Average & 0.463 & \textbf{0.496}\\
		\hline
		& Image$\rightarrow$Text & 0.287 & \textbf{0.312}\\
		NUS-WIDE-10k & Text$\rightarrow$Image & 0.282 & \textbf{0.354}\\
		& Average & 0.285 & \textbf{0.333}\\
		\hline
		& Image$\rightarrow$Text & 0.246 & \textbf{0.304}\\
		Pascal Sentences & Text$\rightarrow$Image & 0.248 & \textbf{0.282}\\
		& Average & 0.247 & \textbf{0.293} \\
		\noalign{\smallskip}\hline
	\end{tabular}
\end{table}

\begin{table}
	\centering
	\caption{MAP scores of our UNCSM approach with or without pretrain stage.}
	\label{table:res_pretrain}       
	\begin{tabular}{lllll}
		\hline\noalign{\smallskip}
		Dataset & Method & Image$\rightarrow$Text & Text$\rightarrow$Image & Average  \\
		\noalign{\smallskip}\hline\noalign{\smallskip}
		Wikipedia & UNCSM without pretrain & 0.335 & 0.297& 0.316 \\
		& \textbf{UNCSM with pretrain} & \textbf{0.360} & \textbf{0.310}& \textbf{0.335} \\
		\hline
		NUS-WIDE & UNCSM without pretrain & 0.481 & 0.468& 0.475\\
		& \textbf{UNCSM with pretrain} & \textbf{0.514} & \textbf{0.477}& \textbf{0.496} \\
		\hline
		NUS-WIDE-10k & UNCSM without pretrain & 0.280 & 0.344& 0.312\\
		& \textbf{UNCSM with pretrain} & \textbf{0.312} & \textbf{0.354}& \textbf{0.333} \\
		\hline
		Pascal Sentences & UNCSM without pretrain & 0.246 & 0.248& 0.247 \\
		& \textbf{UNCSM with pretrain} & \textbf{0.304} & \textbf{0.282}& \textbf{0.293} \\
		\noalign{\smallskip}\hline
	\end{tabular}
\end{table}

As shown in Table \ref{table:res_metric}, we use cosine distance metric and metric network respectively to calculate the cross-media similarity on the shared representation obtained from the two-pathway network. Our metric network achieves significant improvement on 4 datasets compared with sample distance metric such as cosine distance, which proves that our metric network can effectively calculate the complex cross-media similarity. 
Besides, Table \ref{table:res_pretrain} shows the MAP scores of our UNCSM approach with pretrain stage and without pretrain stage, which indicates that pretraining with contrastive loss by modeling the pairwise similar and dissimilar constraints can obviously improve the cross-media retrieval accuracy.

From the above results on 4 datasets in Tables \ref{table:res_wiki} to \ref{table:res_pascal}, we can see that our UNCSM approach can effectively model the cross-media similarity and achieve the best retrieval accuracy among all compared methods. For the traditional methods using linear projection, CFA minimizes the Frobenius norm in the transformed domain which is more suitable for different modalities, making it perform better than CCA. And KCCA can effectively handle the multimedia data with high nonlinearity using kernel functions, which leads it to the best accuracy among the traditional methods. As for the DNN-based methods, Multimodal DBN has the worst performance for it can only model the correlation between different media types by a shallow network, while DCCAE can minimize the reconstruction error additionally which makes it outperform DCCA. And Corr-AE achieves the best results on most of datasets by modeling the reconstruction error as well as correlation loss jointly at the same time. 

However, the performance of the above methods is limited for the fact that they mainly focus on the pairwise similar constraints as well as reconstruction information, and only sample distance metric is used to calculate the cross-media similarity. Compared to the existing methods, our UNCSM approach achieves significant improvement for three reasons: pretraining the two-pathway network with contrastive loss to model the pairwise similar and dissimilar constraints, fine-tuning the network using double triplet similarity loss to preserve the relative similarity for learning the optimized semantic shared representation, and finally embracing more complex similarity functions to calculate the cross-media similarity with the metric network, by modeling the pairwise similar and dissimilar constraints. Our UNCSM approach associates the representation learning with distance metric for further improving the cross-media retrieval accuracy.

\section{Conclusion}

In this paper, a cross-media similarity learning model, UNCSM, has been proposed. This UNCSM model associates the cross-media shared representation learning with distance metric in a unified framework. The UNCSM adopts contrastive loss to pretrain the two-pathway network, and uses double triplet similarity loss for fine-tuning to learn the semantic shared representation for each media type by distance comparisons. And the metric network is employed for effectively calculating the cross-media similarity of the shared representation, by modeling the pairwise similar and dissimilar constraints. 
Compared to the existing methods, our UNCSM approach further improves the cross-media retrieval accuracy by preserving the relative similarity as well as embracing more complex similarity functions at the same time. The experimental results show that our UNCSM approach outperforms 8 state-of-the-art methods on 4 widely-used datasets. As for future work, we attempt to integrate semi-supervised information into our unified metric framework for further boosting the accuracy of cross-media retrieval. 

\begin{acknowledgements}
This work was supported by National Natural Science Foundation of China under Grants 61371128 and 61532005, and National Hi-Tech Research and Development Program of China (863 Program) under Grant 2014AA015102.
\end{acknowledgements}

\bibliographystyle{spbasic}      
\bibliography{ncmt}   


\end{document}